\documentclass{doublecol-new}
\usepackage{graphicx}
\usepackage{natbib,stfloats}
\usepackage{colortbl}
\usepackage{graphics}
\usepackage{amssymb}
\usepackage{amssymb,amsmath}
\usepackage{makeidx}
\makeindex
\usepackage{setspace}
\usepackage{textcomp}
\usepackage{dsfont}
\usepackage{mathrsfs}
\usepackage{wrapfig}
\usepackage{rotating}
\usepackage{multirow}
\usepackage{multicol}
\usepackage[dvipsnames,svgnames,table]{xcolor}
\usepackage{ulem}
\usepackage{hyperref}
\DeclareMathAlphabet{\mathpzc}{OT1}{pzc}{m}{it}
\def\mathbi#1{\textbf{\em #1}}

\theoremstyle{TH}{

}
\theoremstyle{THrm}{

}
\theoremstyle{THhit}{

}
\makeatletter
\def\theequation{\arabic{equation}}
\makeatother
\begin{document}%
\thispagestyle{plain}
\setcounter{page}{1}
\LRH{Nilo Serpa}
\RRH{The Counting of Galaxies from Type Ia Supernovae Rate}
\VOL{1}
\ISSUE{3/4}
\PUBYEAR{2011}
\BottomCatch
\title{The Counting of Galaxies from Type Ia Supernovae Rate}
\authorA{Nilo Serpa}
\affA{Instituto de Ci\^{e}ncias Exatas e Tecnologia,\\
UNIP - Universidade Paulista,\\
SGAS Quadra 913, $s/n^{\underline{o}}$ - Conjunto B - Asa Sul - Bras\'{i}lia - DF, Brasil\\
CEP 70390-130\\ e-mail: nilo@techsolarium.com}
\begin{abstract}
The aim of this paper is to establish a relationship between the number of galaxies and the rate of type Ia supernovae. In addition, present approach prepares a framework to analyze this relationship and its dependence on the redshift ($z$), giving aid to the deepening of studies on radial distribution of galaxies. The work also shows how the counts interrelates to topological aspects regarding the evolution of the universe from $z\sim2$. The relationship was examined apart from morphological considerations.
\end{abstract}
\KEYWORD{type Ia supernova; galaxy counting; host function; redshift}
\begin{bio}
Nilo Serpa is {\it Magister in Scientia}, in Astronomy, from the {\it Universidade do Brasil}, and Master of Business Administration from the {\it Funda\c{c}\~{a}o Getulio Vargas}, Brazil. His {\it Magister} Thesis was about applications of the Lema\^{i}tre-Tolman cosmology with an original approach of weak gravitational lensing in that cosmology. He specializes in Management Process and Information Technology (IT) from the {\it Funda\c{c}\~{a}o Oswaldo Cruz}, Brazil. His training in Physics at the {\it Centro Brasileiro de Pesquisas F\'{i}sicas - CBPF -} ranges from Quantum Mechanics to Supergravity, including Gauge Field Theories, Topological Effects and Supersymmetry. He received his degree of Architect in the year 1981, now having thirty years experience in IT as a Project Manager and Development Manager, and fifteen years experience in Physics as researcher and professor. In the early Eighties, it was pupil of the semiologist Umberto Eco, experience that contributed for his interest on the semantics of the mathematical formalizations. He is Associate Professor of Physics, Software Engineering and Professional Ethics at the {\it Universidade Paulista}, Brazil, and Senior Development Manager at {\it POLITEC Global IT Services}, Brazil. He is Analyst of Training with great experience in e-learning, having worked in collaboration with some of the greatest names in the area of qualification of human resources, such as the Canadian Software Engineer John Franklin Arce. He is also Special Project Manager at the General Coordination of Information and Informatics of the Ministry of Work and Employment, Brazil. Having received a Senior training in Function Point Analysis, he has created a methodology named "Priority Point Analysis" to measure an IT Coordination by the problems it faces. He is author of the book {\it Revers\~{o}es Geopol\'{i}ticas: Geografia, F\'{i}sica e Filosofia na Sociedade Globalizada} (2002). His main works in Physics are "Thermodynamics of Diabetes Mellitus: the Physical Reasons of Obesity and Sedentarismus as Decisive Variables in Predictive Models", "New Lectures on Supergravity", "{\it El-Ni\~{n}o: Influ\^{e}ncia Exterior, Mat\'{e}ria Negra e Caudas Gravitacionais}" and "Modelling the Dynamics of the Work-Employment System with Predator-Prey Interactions". He is member of the review board of the {\it International Journal of Information Technology Project Management} ({\it IJITPM}). His areas of interest include Cosmology, Field Theory, Geopolitics, Econophysics and Information Technology.\break
\end{bio}
\maketitle\vfill\pagebreak
\maketitle
\thispagestyle{empty}
\maketitle
\vspace*{-16pt}
{\it Behold the heavenly space, its stability,
celerity, and quit, finally, to look vile things.
}
\begin{flushright}
{\it Bo\'{e}cio}
\end{flushright}
\section*{The existing type Ia supernovae scenario} 
\addcontentsline{toc}{section}{The existing type Ia supernovae scenario}
\subsection*{The supernovae counting}
\addcontentsline{toc}{subsection}{The supernovae counting}
Type Ia supernovae are among the biggest thermonuclear explosions in the Universe. During the two last decades, they gained great importance in cosmology as benchmarks to measure distances due to their regular behavior like standard candles and because of their essential role in the search for decisive facts to corroborate the accelerated expansion of the Universe. At first, there was a belief that type Ia supernovae (SNe) originate only from a carbon-oxygen white dwarf (WD) in a binary system, where the dwarf companion is a red giant. Nowadays, two possible progenitor models are discussed, that is, a) one starting from the classical hypothesis that the WD, accreting from its main sequence or red giant companion, grows in mass until it reaches the critical Chandrasekhar limit, and b) other starting from the hypothesis of a close double WD system merging after orbital shrinking due to the emission of gravitational wave radiation. In both cases, the nature and evolution of the binary system remains a puzzle. The time elapsed from the birth of the binary system to the SN explosion, called "delay time", spans from tens of millions of years to ten billion years or even more. As a consequence, the counting of SNe Ia (or the rate of generation) reveals the star formation history of a galaxy accordingly the distribution of the delay times. The analysis of the SNe Ia rate as a function of redshift and the inquiry whether their properties evolve with redshift would be powerful tools to aid the investigation of the nature of the galaxy environments and their connections with the rate of galaxy formation itself.
Type Ia supernovae have been key pieces in studies on the true nature of the distribution of matter in the universe. The large number of uncertainties referred to the immensity of the distances and time scales leaded to alternatives to the more accepted interpretation of the data, which was taken in accordance with the homogeneous Friedmann-Lema\^{i}tre-Robertson-Walker (FLRW) cosmology. In this way, a simple and realistic option is provided by the inhomogeneous Lema\^{i}tre-Tolman (LT) cosmology, applied and studied by several authors. For instance, there is an article by D. Garfinkle in which an inhomogeneous LT spacetime is treated in an understandable and very pragmatic way to modeling type Ia supernovae data \citep{7}. By its simplicity and clarity, I consider this study as a fundamental starting point for further research on inhomogeneous cosmology. Also based on the adoption of a LT cosmology, K. Bolejko analyzes whether the modeling of inhomogeneous distribution of matter with no cosmological constant is realistic or not, taking as object of investigation the reduction of brightness of type Ia supernovae \citep{2}. 
The SNe rate at a given redshift is usually computed as the ratio between the number of known SNe Ia and the control time of the monitored galaxies at that redshift. One may object that statistics would be a poor tool to model a tracer for the rates both at low redshift, due to the difficulty of sampling large volumes, and at high redshift, due to the difficulty of detecting and typing faint SNe. However they are still seen as rare and transient events, thousands of SNe Ia are now discovered per year \citep{6}, and the deep sky observations with suitable time interval required to enlarge the number of real SNe candidates are more and more attainable as technology advances. Thus, the more the range of $z$ is wide, the more will be the reliability on the proposed $z$-dependence of the rate. 
The main motivations of this work were a)- to relate quantitatively the countings of galaxies and type Ia supernovae, trying to bring new elements to contrast conventional searching on radial distribution of galaxies, b)- to find and discuss possible evolutionary topological properties of the Universe from $z\sim2$ based on that quantitative relation, and c)- to question weather the eventual existence of evolutionary "jumps" \hspace{0.6 mm} in the Universe would provoke relevant anisotropies on CMB. The central idea is to define a relation between the number of galaxies and the SNe rate by redshift, since we know that the energy deposition from supernovae into the environment is substantial to star formation and even to galaxy formation. In addition, discounting all troubled observational biases, such as that caused by the interstellar dust, which produces more extinction of the high redshift supernovae, distinct rates for different high redshifts could result from the evolution of progenitor systems by variation in the mass, composition, and metallicity of type Ia supernovae. The test of that hypothesis requires accurate measurements of SN rates at various cosmic epochs. Beside, to minimize the uncertainty in the estimates of SN rates, a statistically significant SNe sample is necessary. 
The chemical enrichment of the galaxies by the returns of synthesized elements from the explosive burning processes of supernovae is on its own a strong reason to investigate the explosion rates by redshifts and this is a matter of many works. The literature on galaxy and supernova countings is copious. At first, I distinguish the works on supernova rate by Pain {\it et al} (1996), Cappellaro {\it et al} (1999) and from the EROS Collaboration (2000), followed by Blanc {\it et al} (2004) and culminating with Sharon {\it et al} (2010). In particular, about galaxy number counts and clusters of galaxies we have fine works from Campos (1995), Metcalfe {\it et al} (2000), Grahan {\it et al} (2008) and Labini {\it et al} (2009). Another fundamental study is the classical paper of Lilly {\it et al} (1995).
As low redshifts are the regions where the effective work to understanding SNe Ia must be done, Cappellaro {\it et al} proposed to estimate the SNe rate in the local Universe as a function of both galaxy morphological type and colors. On the other hand, the EROS Collaboration, searching nearby supernovae from $z \sim 0.02$ to $0.2$, took the explosion rate of $\mathcal{N}_{obs}$ observed type Ia supernovae by the expression\\
\begin{equation}
\mathcal{R}=\mathcal{N}_{obs}/\mathcal{S},
\end{equation}
\vspace*{10pt}
where\\
\begin{equation}
\mathcal{S} = \sum\limits_{gal{\rm{ }}\;i} {L_i } \int_{ - \infty }^{ + \infty } {\epsilon _i \left( {t,z_i } \right)dt}, 
\end{equation}
with $\epsilon _i \left( {t,z_i } \right)$ as the efficiency to detect in galaxy $i$ a type Ia supernova whose maximum occurs at time $t$ in the supernova rest frame. Galaxy $i$ has a blue luminosity $L_i$, and the integral $\int_{ - \infty }^{ + \infty } {\epsilon _i \left( {t,z_i } \right)dt}$ is called its "control time". For the EROS simulations, the sum $\mathcal{S}$ was proved to be equal to $9.09 \times 10^{12} h^{-2} L_{\odot B}\; yr$ and the error is statistical and usually given at a 68$\%$ confidence level. Further, also according to EROS collaboration, there must be a cut on the galaxy magnitude, so that we may ensure that the host galaxy for a type Ia supernova is classified as such during the visual scanning. For example, at $z = 0.1$, the cut corresponds to an absolute magnitude of $M_{gal} \sim -20. + 5 log(H_0/60 kms^{-1}Mpc^{-1})$ for the host galaxy. 
Possible samples of supernovae are available in some good catalogs. An interesting sample from ESSENCE Supernova Survey (Miknaits+, 2007) is shown in Table 1 with probabilities for the real type Ia supernovae candidates. In this paper, I present a general form to predicting the number of galaxies associated to the type Ia supernova explosion rate. I assumed that the expected rate at a given red shift is proportional to the amount of galaxies at the redshift. In this approach there are no concerns about morphology but purely statistical considerations. Since in this study no attempt was made to derive rates for different galaxy types and the individual SN rate was implicitly assumed depending on host galaxy luminosity, the predicted number of galaxies is such that the rate is achieved by the presence of enough galaxies with appropriated luminosity. 
\begin{center}
\tiny{{\bf Table 1: type Ia supernovae from ESSENCE Supernova Survey}
\begin{tabular}{lrrr|lrrr}
$sn$ & $p$ & $z_{g}$ & $z_{sn}$ & $sn$ & $p$ & $z_{g}$ & $z_{sn}$\\ 
{\it 2002iu} & 100.0 & ---& 0.115 & {\it 2003ji} & 96.9 & ---& 0.211 \\
{\it 2002iv} & 98.3 & 0.231 & 0.226 & {\it 2003jq} & 98.3 & ---& 0.156 \\
{\it 2002jq} & 81.4 & ---& 0.474 & {\it 2003jw} & 100.0 & 0.296 & 0.309 \\
{\it 2002iy} & 82.4 & 0.587 & 0.59 & {\it 2003jy} & 100.0 & 0.339 & 0.342 \\
{\it 2002iz} & 98.6 & 0.428 & 0.426 & {\it 2003kk} & 100.0 & 0.164 & 0.159 \\
{\it 2002ja} & 100.0 & ---& 0.329 & {\it 2003kl} & 100.0 & 0.335 & 0.332 \\
{\it 2002jb} & 100.0 & ---& 0.258 & {\it 2003km} & 100.0 & ---& 0.469 \\
{\it 2002jr} & 100.0 & ---& 0.425 & {\it 2003kn} & 100.0 & 0.244 & 0.239 \\
{\it 2002jc} & 65.7 & ---& 0.54 & {\it 2003ko} & 99.2 & 0.36 & 0.352 \\
{\it 2002js} & 100.0 & ---& 0.55 & {\it 2003kt} & 100.0 & ---& 0.612 \\
{\it 2002jd} & 96.6 & ---& 0.318 & {\it 2003kq} & 100.0 & 0.606 & 0.631 \\
{\it 2002jt} & 100.0 & ---& 0.382 & {\it 2003kp} & 100.0 & ---& 0.645 \\
{\it 2002ju} & 100.0 & 0.348 & 0.35 & {\it 2003kr} & 100.0 & 0.427 & 0.429 \\ 
{\it 2002jw} & 100.0 & 0.357 & 0.362 & {\it 2003ks} & 98.6 & ---& 0.497 
\\
---& 100.0 & 0.399 & 0.4 & {\it 2003ku} & ---& ---& --- \\
{\it 2003jo} & 96.0 & 0.524 & 0.531 & {\it 2003kv} & ---& ---& --- \\
{\it 2003jj} & 95.0 & 0.583 & 0.583 & {\it 2003lh} & 100.0 & ---& 0.539 \\
{\it 2003jn} & 100.0 & ---& 0.333 & {\it 2003le} & 100.0 & ---& 0.561 \\
{\it 2003jm} & 100.0 & 0.522 & 0.519 & {\it 2003lf} & 100.0 & ---& 0.41 \\
{\it 2003jv} & 100.0 & 0.405 & 0.401 & {\it 2003lm} & 100.0 & 0.408 & 0.412 \\
{\it 2003ju} & 100.0 & ---& 0.205 & {\it 2003ll} & 100.0 & 0.596 & 0.599 \\
{\it 2003jr} & 100.0 & 0.34 & 0.337 & {\it 2003lk} & 33.3 & 0.442 & --- \\
{\it 2003jl} & 100.0 & 0.429 & 0.436 & {\it 2003ln} & 100.0 & ---& 0.619 \\
{\it 2003js} & 93.4 & 0.363 & 0.36 & {\it 2003lj} & 100.0 & 0.417 & 0.422 \\
{\it 2003jt} & 100.0 & ---& 0.436 & ---& ---& ---& --- \\ 
\end{tabular} 
} 
\end{center}
\section*{Frameworks to counting galaxies from SNe explosion rate}
\addcontentsline{toc}{section}{Frameworks to counting galaxies from SNe explosion rate}
\subsection*{The classical galaxy counting}
\addcontentsline{toc}{subsection}{The galaxy counting}
The numerical differential counting of objects (galaxies or clusters, depending on the 
numerical density and the source masses), or solely differential counting, is defined in terms of the volumn as
\begin{equation}
dN = n(t)dV,
\end{equation}
where $n(t)$ is the numerical density ($V^{-1}$) and $dV$ the volumn element. This last is givem by 
\[
dV = \sqrt g {\rm{ }}drd\theta d\phi = \frac{{a^3 r^2 dr}}{{\sqrt {1 - kr^2 } }}\sin \theta {\rm{ }}d\theta d\phi, \hspace{14 mm} (3.a) 
\]
\[
dV = \frac{{4\pi a^3 r^2 dr}}{{\sqrt {1 - kr^2 } }}, \hspace{50 mm} (3.b)
\]
where $a$ is the scale factor or expansion rate of the Universe, $r$ is the commoving radius, $k$ is a constant of curvature and $g$ is the determinant of the Robertson-Walker metric tensor. From these expressions we deduce that $dV$ is a spherical shell with radius $ar$ and thickness $(1-kr^2)^{-1/2}adr$.
In terms of redshift, the connection between the forecasted behavior of the differential counting and that builded from the observations is given by
\begin{equation}
\left[ {\frac{{dN}}{{dz}}} \right]_{obs} = \frac{{\psi (z)}}{{n_c (z)}}\frac{{dN}}{{dz}},
\end{equation}
where $\psi (z)$ is the selection function applied on the required catalog and $n_c (z)$ is the radial numerical density in terms of the commoving volumn. A more complete approach on this subject may be found, for instance, in Iribarrem (2009).
\subsection*{First approach}
\addcontentsline{toc}{subsection}{First approximation of the host function}
Let us take the EROS surveying; beside, let us assume that the star formation activity is steady and continuous for most of galaxies \citep{17}, and for any redshift, starting from the requirement that one SN must be located at one host galaxy, we have a SN rate and a differential galaxy counting related by
\begin{equation}
\mathcal{R} = \mathbb{T}(z)\left[\frac{{dN}}{{dz}}\right]_{obs} = \frac{\mathcal{N}_{obs}}{\mathcal{S}},
\end{equation}
which means there is a relation between the number of galaxies and the number of type Ia supernovae, strongly depending on $z$, with $\mathbb{T}(z)$ being a certain function of the redshift to be defined ahead. Since $dN/dz$ is independent of the cosmology, getting the differential counting for a certain catalog we may choose the cosmological model further. From equations (1) and (2) we have
\begin{equation}
dN_{obs}= \mathcal{N}_{obs}\frac{{dz}}{{\mathbb{T}(z)\mathcal{S}}},
\end{equation}
\begin{equation}
N_{obs} = \int \mathcal{N}_{obs}{\frac{{dz}}{{\mathbb{T}(z)\mathcal{S}}}}, 
\end{equation}
\begin{equation}
N_{obs} = \int {\frac{{\mathcal{N}_{obs}dz}}{{\mathbb{T}(z)\sum\limits_{gal{\rm{ }}\;i} {L_i } \int_{ - \infty }^{ + \infty } {\epsilon _i \left( {t,z_i } \right)dt} }}}. 
\end{equation}
As $z_i$ refers to the $i$ redshifts of a certain observed galaxy sample, the differential $dz$ takes account only of the integration of $1/\mathbb{T}(z)$. For $\mathcal{S}= 9.09 \times 10^{12} h^{-2} L_{\odot B}\; yr$, equation (8) reduces to
\begin{equation}
N_{obs} = \int {\frac{{\mathcal{N}_{obs}dz}}{{\mathbb{T}(z)\times 9.09 \times 10^{12} h^{-2} L_{\odot B}\; yr }}}. 
\end{equation}
According to EROS, there were surveyed 80 square degrees, comprising eight detected supernovae, four of which spectroscopically identified as type Ia supernovae. Thus,
\begin{equation}
N_{obs} = \frac{4}{9.09 \times 10^{12} h^{-2} L_{\odot B}\; yr} \int {\frac{{dz}}{{\mathbb{T}(z) }}}. 
\end{equation}
\subsection*{The general host function}
\addcontentsline{toc}{subsection}{The general host function}
On the other hand, accordingly Pain {\it et al} (1996) in an interesting study on the expected number of supernovae as a function of $z$, the observed SN rate is also given by
\begin{equation}
\mathcal{R} = \frac{{r_{sn} }}{{(1 + z)}},
\end{equation}
where $r_{sn}$ is the rate in the rest-frame of the supernovae. Thus, we have 
\begin{equation}
\mathbb{T}(z)\left[\frac{{dN}}{{dz}}\right]_{obs} = \frac{\mathcal{N}_{obs}}{\mathcal{S}},
\end{equation} 
\begin{equation}
\frac{{r_{sn} }}{{(1 + z)}} = \frac{\mathcal{N}_{obs}}{\mathcal{S}}.
\end{equation}
From equations (12) and (13) we get
\begin{equation}
\left[\frac{{dN}}{{dz}}\right]_{obs} = 1/\mathbb{T}(z) \frac{{r_{sn} }}{{(1 + z)}},
\end{equation}
\begin{equation}
N_{obs}= \int \frac{{r_{sn}dz }}{{\mathbb{T}(z)(1 + z)}}.
\end{equation}
Now we equal equations (8) and (15), so that
\begin{equation}
\int {\frac{{\mathcal{N}_{obs}dz}}{{\mathbb{T}(z)\sum\limits_{gal{\rm{ }}\;i} {L_i } \int_{ - \infty }^{ + \infty } {\epsilon _i \left( {t,z_i } \right)dt} }}}=\int \frac{{r_{sn}dz }}{{\mathbb{T}(z)(1 + z)}},
\end{equation}
\begin{equation}
\frac{{\mathcal{N}_{obs}}}{{\mathbb{T}(z)\sum\limits_{gal{\rm{ }}\;i} {L_i } \int_{ - \infty }^{ + \infty } {\epsilon _i \left( {t,z_i } \right)dt} }}= \frac{{r_{sn}}}{{\mathbb{T}(z)(1 + z)}},
\end{equation}
\begin{equation}
\mathcal{N}_{obs}= \frac{{r_{sn}}}{{(1 + z)}}\sum\limits_{gal{\rm{ }}\;i} {L_i } \int_{ - \infty }^{ + \infty } {\epsilon _i \left( {t,z_i } \right)dt},
\end{equation}
which is precisely equation (13). The function $\xi=r_{sn}/\mathbb{T}(z)(1+z)$ is called "host function" and it brings the cosmological phase signature at a given $z$ for a certain host galaxy. The signatures are positive till $z\sim0.5$ (figure 1). The change in the signal of the host function is consistent with the change from a decelerating Universe (matter dominated phase) up until about 5 billion years ago (around $z=0.5$) to the current accelerating stage \citep{11}. At certain $z$, the integration of the host function gives the expected number of galaxies related to the expected rate of supernovae; so, function $\mathbb{T}(z)$ acts upon the derivative of the number of galaxies at that $z$ to give the rate of supernovae for the respective $z$-shell. As very rare events, statistically speaking, we may suppose from the viewpoint of the observer one supernova to one galaxy, so that the rate of supernovae is in sum closely the rate of host galaxies.
\begin{figure} [h]
\includegraphics[scale=0.42]{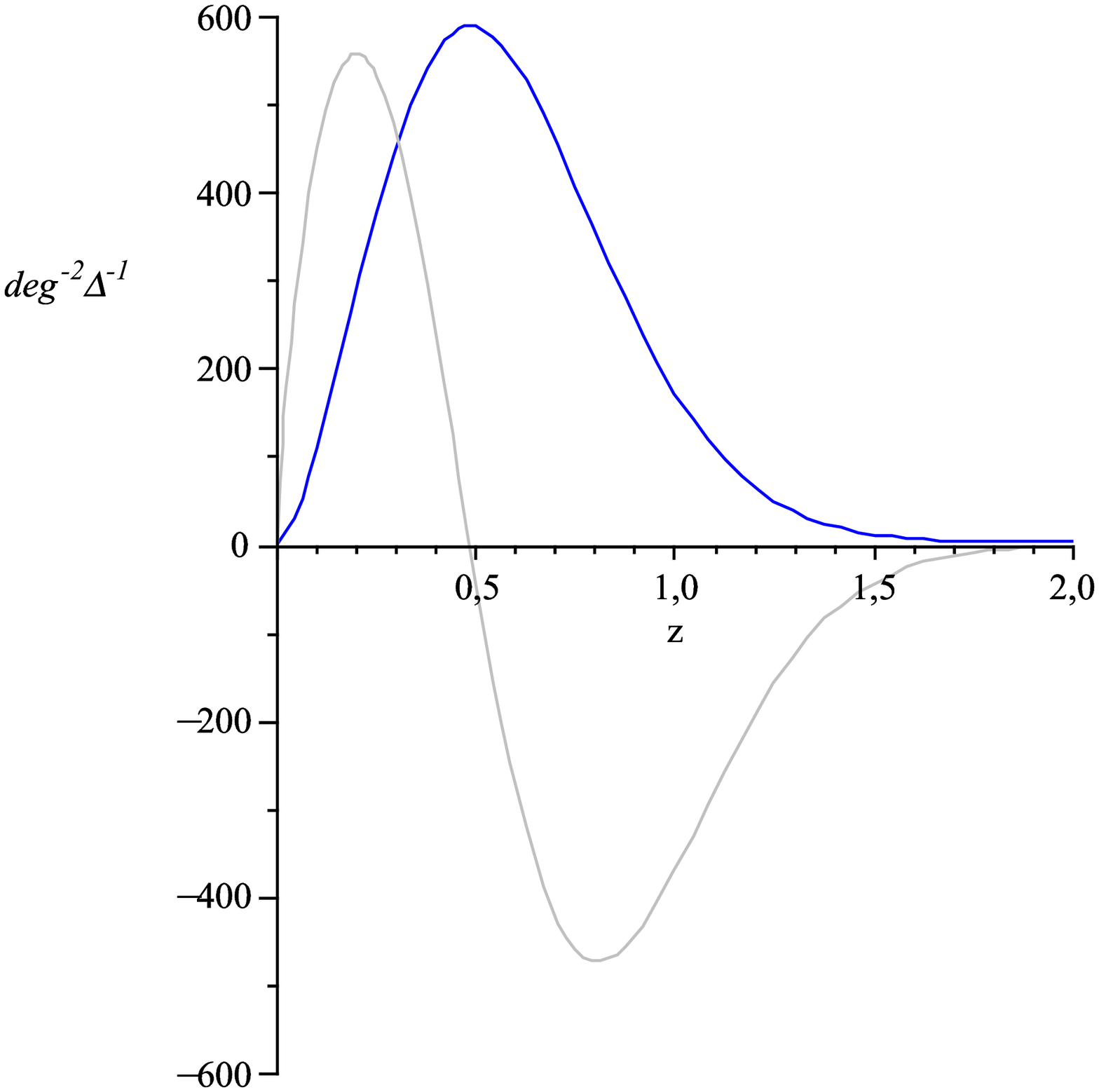}
\\
\small {Figure 1: {\it evolution of host function $\xi$ (grey) and number of galaxies $N_{obs}$ (blue) until $z=2$.}} 
\end{figure}
\begin{figure} [h]
\includegraphics[scale=0.42]{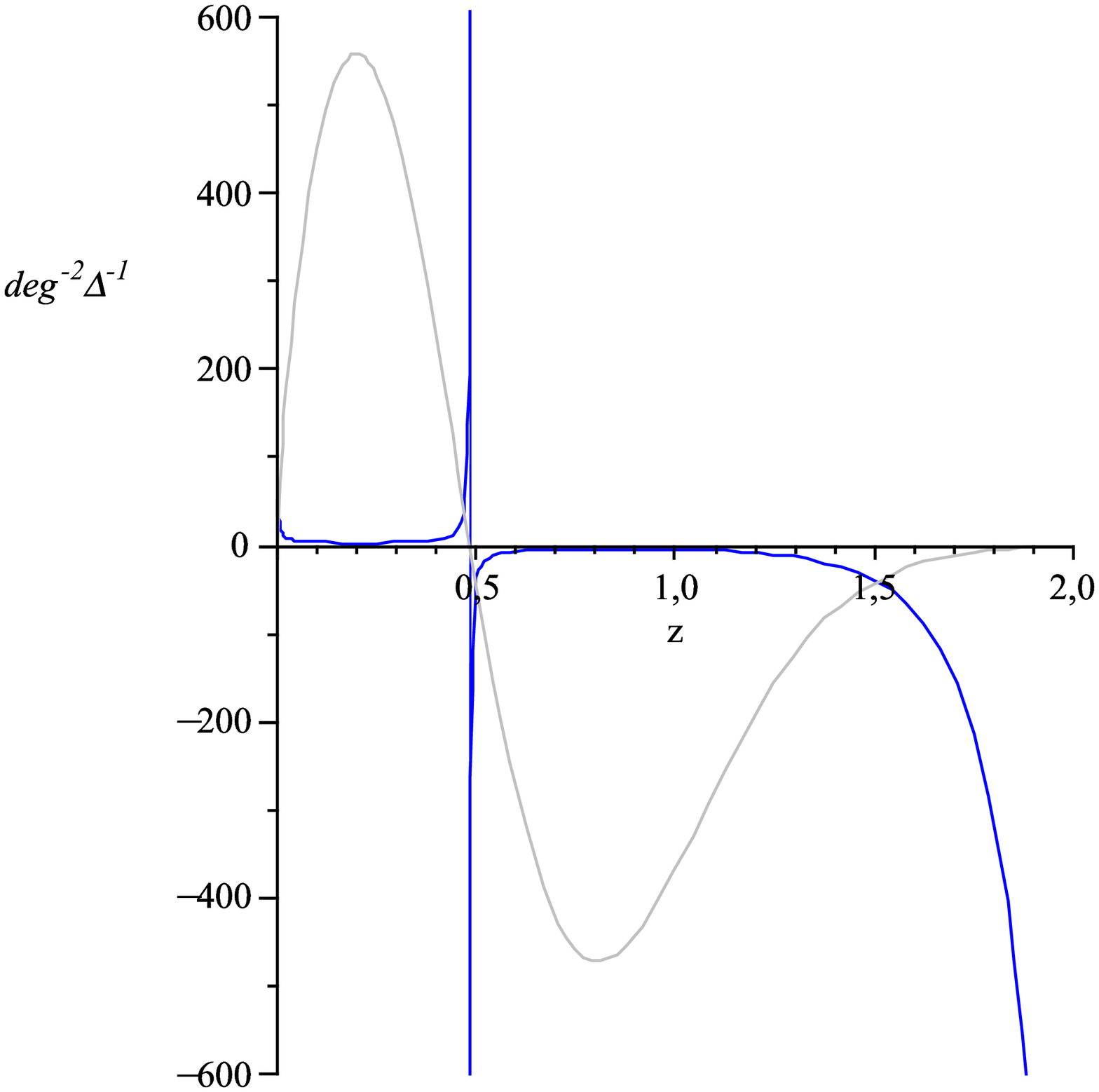}
\\
\small {Figure 2: {\it evolution of $\mathbb{T}$ (blue) and $\xi$ (grey) until $z=2$.}} 
\end{figure}
However, as pointed out by Broadhurst, Taylor and Peacock (1994), we may assume some cosmological premises to fit the data of redshift surveys, remembering that the representative function of the number of galaxies $N_{obs}$ is model-independent. The numerical results for $N_{obs}$ were obtained from the given luminosity function and these authors worked at the faint limit for spectroscopy in the $R$ band, in which $R=22.5$. They adopted for this case the expression
\begin{equation}
N_{obs}= 11.7 z^{1.63} exp[-(z/0.51)^{1.79}],
\end{equation}
where $N_{obs}$ refers to the probability distribution for redshift; I used that to infering the host function precisely for this case. Equations (15) and (16) furnish
\begin{equation}
11.7 z^{1.63} exp[-(z/0.51)^{1.79}] = \int \frac{{r_{sn}dz }}{{\mathbb{T}(z)(1 + z)}},
\end{equation}
from which we gain
\begin{equation}
\frac{{r_{sn}}}{{\mathbb{T}(z)(1 + z)}}= 19.07 z^{0.63}(1-3.66 z^{1.79}) e^{-z^{1.79}/0.2996}.
\end{equation}
The last equation gives
\begin{equation}
\mathbb{T}(z)= \frac{{r_{sn}e^{z^{1.79}/0.2996}}}{{19.07 z^{0.63}(1-3.66 z^{1.79})(1 + z)}}.
\end{equation}
Substituting this result, the integration in equation (15) furnishes
\[ 
N_{obs} =\int _{0}^{z} \!-{\frac {1907}{1250}}\,{e^{-{\frac 
{104305741}{31250000}}\,{{\it z}}^{{\frac {179}{100}}}}}{{\it z}
}^{{\frac {21}{50}}} \times
\]
\begin{equation}
\left( -50\,{{\it z}}^{{\frac {21}{100}}}+183
\,{{\it z}}^{2} \right) {d{\it z}}+1 .
\end{equation}
We note that $\mathbb{T}$ has a discontinuous structure with different limits at a particular point around $z\sim 0.5$ depending on whether this point is approached from the positive or negative directions (Figure 2). In addition, the two "foils" of $\mathbb{T}$ are not symmetric. In fact, one may define a map $\mathcal{M}_f:\mathbb{T}(z)\Longrightarrow\xi$ to cover $\mathbb{T}$, that is, to bypass the discontinuity. This covering sounds artificial, and I defend the opinion that the $\mathbb{T}(z)$-jump at $z\sim 0.5$ must be a real break in the cosmic evolution related to the harsh phase transition toward an accelerating Universe (the hardness to understand or to explain the physics of the jump is not acceptable as an argument to bypass discontinuity). The discontinuous structure $\mathbb{T}$ is featured by one singularity. So, since limits are topological constructions, that is, manipulations with limits are based on the topology of number spaces, let $\mathord{\buildrel{\lower3pt\hbox{$\scriptscriptstyle\smile$}} 
\over z}$ be a singular point around $z\sim 0.5$; also let $\Im : = (Z,T)$ be a topological space. We say that $\mathord{\buildrel{\lower3pt\hbox{$\scriptscriptstyle\smile$}} 
\over z}$ is an {\it isolated point} of $\Im : = (Z,T)\Leftarrow:\Longrightarrow\lbrace\mathord{\buildrel{\lower3pt\hbox{$\scriptscriptstyle\smile$}} 
\over z}\rbrace\in T$, such that all other points are {\it limit points} enclosed in the set $Lim(\Im):=\lbrace z\in Z\mid{z} \notin T \rbrace$. In this case, $\Im$ is said to be a {\it singular space}. Thus, defining the function $\mathbb{T}(z)$ we are building a map $\mathcal{M}_\Im:\Im\Longrightarrow\Im'$ that carries topological space $\Im$ into topological space $\Im'$. Both spaces are subspaces of $\mathbb{R}$. The intervals of $z$ are $0\leq z$ $\leq$ $\mathord{\buildrel{\lower3pt\hbox{$\scriptscriptstyle\smile$}} 
\over z}$ and $\mathord{\buildrel{\lower3pt\hbox{$\scriptscriptstyle\smile$}} 
\over z}$ $< z \leq 2$; the union of the two non-symmetric disjoint foils gives $\mathbb{T}$. Because of its singularity, which means that we need to manipulate limits, and by the fact that it is not depending on the metric (cosmology), function $\mathbb{T}(z)$ was called "Topological Evolution from High Redshift Astronomy" (TEHRA). This function may bring some interesting additional insights to relativistic cosmology; as partial differential equations, Einstein's equations describe only local properties of spacetime and do not give any information about the global structure (the topology) of the spacetime. 
In physics, topological effects connected to non-genetic holes or harsh breakouts on manifolds are generally expensive in energy. This cost in energy must be detectable such as anisotropies on CMB. Nevertheless, in present state of knowledge it is very difficult to decide if the TEHRA singularity (TS) originated some kind of anisotropy on CMB. Such anisotropy, if detected, would be an indirect evidence of TS after an accurate analysis of all possibilities. Recently, I suggested to my colleagues a project to investigate exhaustively the Sunyaev-Zel'dovich effect from $z<0.5$ and $z>0.5$ trying to detect some additional harsh distortion in the expected gained energy of the electrons that, crossing the core of a galaxy cluster, are Thomson scattered distorting the CMB blackbody spectrum. The essential point to be investigated is the way one might decide on the origins of certain anisotropy in the cosmic microwave background. For instance, the spectral shapes of the anisotropy generated by the gravitational lensing are identical to those generated by kinematic Sunyaev-Zel'dovich effect.
\section*{Conclusion}
\addcontentsline{toc}{section}{Final remarks}
Present work explained what I called "host function" as an attempt to connect type Ia supernovae rate and the counting of galaxies in accordance to the observational data. The central idea is that for a certain rate of supernovae there are sufficient host galaxies to warrant the rate. There is a wide field for detailed research in this way to discussing the host function. For instance, some possible physical consequences of evolutionary topological features pointed by the model were explained, suggesting further investigations and tests on CMB to look for relics from the singularity predicted. In addition, however the counting of galaxies is independent on cosmology, it would be particularly interesting to take into account the inhomogeneous Lema\^{i}tre-Tolman and the Friedmann-Lema\^{i}tre-Robertson-Walker cosmologies to compare the distribution of galaxies deduced from the above theoretical approach with that predicted by both cosmological models. 
\begin{center}
$$
$$
$\diamondsuit\diamondsuit\diamondsuit$
\end{center}
\markright{\bfseries } 

\printindex
\begin{center}
$\diamondsuit\diamondsuit\diamondsuit$
\end{center}
\end{document}